\documentclass[sn-standardnature,default,iicol]{sn-jnl}%
\usepackage{siunitx}
\usepackage{pdfpages}

\jyear{2021}%

\theoremstyle{thmstyleone}%

\theoremstyle{thmstyletwo}%

\theoremstyle{thmstylethree}%

\unnumbered

\usepackage[switch]{lineno} 

\begin{document}

\title[$\quad$]{Spatiotemporal single-photon Airy bullets}

\author[1,2]{\fnm{Jianmin} \sur{Wang}}

\author[1,2]{\fnm{Ying} \sur{Zuo}}

\author[1,2]{\fnm{Xingchang} \sur{Wang}}

\author[3]{\fnm{Demetrios N.} \sur{Christodoulides}}

\author*[1,2]{\fnm{Georgios A.} \sur{Siviloglou}}\email{siviloglouga@sustech.edu.cn}

\author*[1,2]{\fnm{J. F.} \sur{Chen}}\email{chenjf@sustech.edu.cn}

\affil[1]{\orgdiv{Shenzhen Institute for Quantum Science and Engineering and Department of Physics}, \orgname{Southern University of Science and Technology}, \orgaddress{\city{Shenzhen}, \postcode{518055}, \state{Guangdong}, \country{China}}}

\affil[2]{\orgdiv{Guangdong Provincial Key Laboratory of Quantum Science and Engineering}, \orgname{Southern University of Science and Technology}, \orgaddress{ \city{Shenzhen}, \postcode{518055}, \state{Guangdong}, \country{China}}}

\affil[3]{\orgdiv{Ming Hsieh Department of Electrical and Computer Engineering}, \orgname{University of Southern California}, \orgaddress{ \city{Los Angeles}, \postcode{90089}, \state{CA}, \country{USA}}}

\abstract{Uninhibited control of the complex spatiotemporal quantum wavefunction of a single photon has so far remained elusive even though it can dramatically increase the encoding flexibility and thus the information capacity of a photonic quantum link. By fusing temporal waveform generation in a cold atomic ensemble and spatial single-photon shaping, we hereby demonstrate for the first time complete spatiotemporal control of a propagation invariant (2+1)D Airy single-photon optical bullet. These correlated photons are not only self-accelerating and impervious to spreading as their classical counterparts, but can be concealed and revealed in the presence of strong classical light noise. Our  methodology allows one to synthesize in a robust and versatile manner arbitrary quantum nonspreading spatiotemporal light bullets and in this respect could  have  ramifications  in  a  broad  range  of  applications  such  as   quantum imaging, long-distance quantum communications, and multidimensional information encoding.}

{

}

\maketitle

\maketitle

In the quest for boosting the information capacity and security of tomorrow’s quantum communication links, considerable effort has been made towards harnessing the properties of entangled flying photons within their available degrees of freedom, e.g. frequency, polarization, and orbital angular momentum, to mention a few~\cite{Lu2020,Torres2007,Kim2014,Kumar2014,Kues2017,Srinivasan2011,Peer2005, Boucher2021, Erhard2020,  Ansari2018, Fleischer2018, Chrapkiewicz2016}. Remarkably, and in spite of this intense activity, the spatial and temporal degrees of freedom---perhaps the most archetypical ones---have not been simultaneously exploited in generating robust propagation invariant information-carrying wave packets~\cite{DiTrapani2012, Ren2021}. 

One of the most prominent members of such nonspreading wavefronts is that associated with self-bending optical fields-like the Airy beam. As opposed to any other class of nondiffracting waves that relies on conical superposition~\cite{Eberly1987}, these intriguing self-healing waves tend to freely propagate with minimal expansion while notably their intensity features move in a self-similar manner along curved trajectories. After their experimental observation, Airy beams~\cite{Siviloglou2007a} have found numerous applications in optical and electron microscopies~\cite{Vettenburg2014, Wang2020, VolochBloch2013}, plasma generation~\cite{Polynkin2009, Tzortzakis2010}, ultrafast optics~\cite{Chong2010} and hot atomic vapors~\cite{Wei2014}, as well as in microparticle manipulation~\cite{Baumgartl2008}. Even though their extraordinary attributes have been extensively investigated within the realm of classical optics~\cite{Efremidis2019}, the same is not true for their purely quantum counterparts that have thus far remained relatively unexplored. 

In recent studies, nonclassical Airy beams have been demonstrated by manipulating their spatial degrees of freedom through spontaneous parametric down-conversion (SPDC) in nonlinear crystals~\cite{Maruca2018,Lib2020, Li2021}. Typically, in such arrangements, the photons are wideband with sub-picosecond coherence times and their spectral brightness is rather limited. On the other hand, entangled paired photons generated from cold ensembles of identical atoms~\cite{Du2008a} are exceptionally bright since the spectrum of the emitted photons can be several times narrower than the atomic natural linewidth which is typically only a few MHz. This makes photons produced from cold atoms ideal for carrying time-multiplexed quantum information over long distances. Their inherent endurance to decoherence is complemented by their unsurpassed spectral purity and the fact that the atoms in cold ensembles act universally as identical emitters~\cite{Yang2016NP,Duan2020PRL,Chen2016}.

In the temporal domain, an Airy-modulated photon waveform happens to be the only nontrivial (not sinusoidal) nonspreading wave packet when propagating in a dispersive environment. Evidently, merging Airy quantum wave packets in space and time can be used to synthesize spatiotemporal photon bullets~\cite{Silberberg1990} that can traverse free space spatially undistorted while also being impervious to temporal dispersion effects, such as in atomic media with steep spectral features. Generating a nonclassical photon bullet is by no means a straightforward task given that the photon statistics always reflect the quantum/classical nature of the optical source, even when dimmed at extremely low levels. Therefore, realizing optical bullets~\cite{Chong2010} in quantum settings will require altogether new strategies for single-photon manipulation. 

In this work, we demonstrate for the first time nonspreading, nondiffracting and self-accelerating spatiotemporal Airy bullets in the quantum realm. This is achieved using heralded single photons with subnatural linewidth and thus long coherence time, produced in an elongated atomic medium by transferring the spatial degree of freedom into the temporal domain (space-time morphing). We show that these single-photon self-accelerating and propagation-invariant wave packets can be retrieved even when embedded in highly noisy environments. As such, they can find applications in long-distance quantum communication links and quantum imaging.  
 
{

}
\section{Results}\label{sec2}

\begin{figure*}[htbp]
  \centering
  \includegraphics[width=1.0\linewidth]{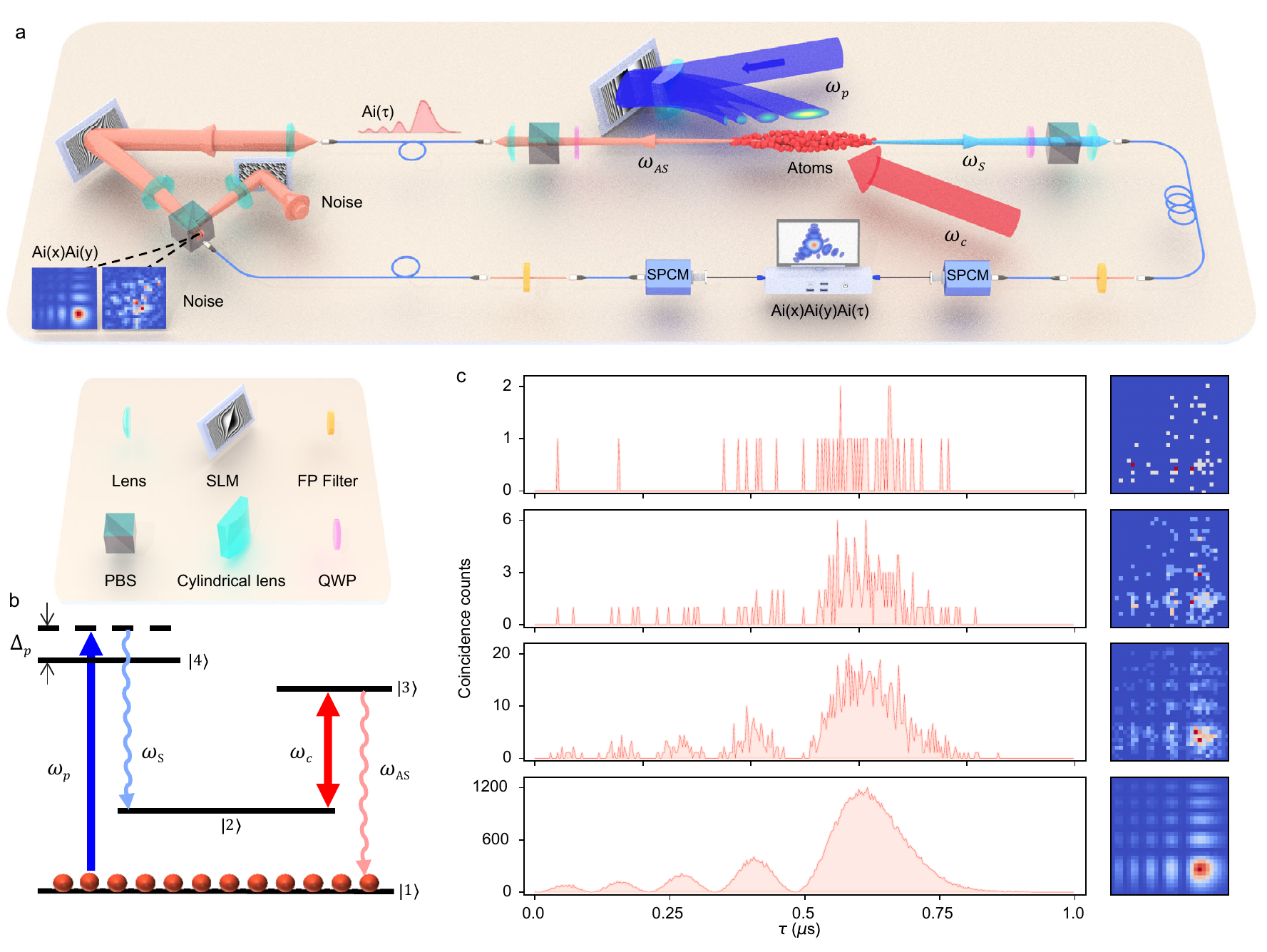}
\caption{$\mathbf{\vert}$\textbf{ Experimental scheme for the observation of spatiotemporal Airy biphotons. }\textbf{a,} Optical setup for the generation of quantum Airy wave packets. Two classical counter-propagating beams with frequencies $\omega_p$ (Airy) and $\omega_c$ (Gaussian) generate Stokes, $\omega_S$, and anti-Stokes, $\omega_{AS}$, entangled photons via spontaneous four-wave mixing in a cold atomic ensemble of rubidium. A space-to-time mapping generates nonclassical temporal Airy biphotons $Ai\left(\tau\right)$, where $\tau$ is the time delay between the arrival of the trigger $\omega_S$ and heralded $\omega_{AS}$ photons. The transverse profile of the $\omega_{AS}$ photons is subsequently shaped into a nondiffracting Airy wave packet to form a three-dimensional spatiotemporal photon, $Ai\left(x\right)Ai\left(y\right)Ai\left(\tau\right)$. Single-photon counting modules are used to register the photons after they are spectrally filtered by Fabry-P\'{e}rot (FP) cavities. A classical speckled light pattern with frequency $\omega_{AS}$ is superimposed for correlated photon concealing. \textbf{b,} Double-$\Lambda$ atomic configuration for spontaneous four-wave mixing. The atoms are initially prepared in the ground state $\vert 1\rangle$. The $\SI{780}{\nano\meter}$ 1D-Airy pump beam, $\omega_p$, is $\Delta_p = \SI{120}{\MHz}$ blue-detuned from the transition $\vert 1 \rangle \rightarrow \vert 4 \rangle$ while the $\SI{795}{\nano\meter}$ coupling beam, $\omega_c$, is on resonance with the $\vert 2 \rangle \rightarrow \vert 3 \rangle$ transition. The twin photons $\omega_{S}$ ($\SI{780}{\nano\meter}$) and $\omega_{AS}$ ($\SI{795}{\nano\meter}$) are spontaneously emitted.
\textbf{c,} Simulated formation of spatiotemporal quantum wave packets photon-by-photon. On the (left), the temporal biphoton waveform takes the distinct Airy profile as the accumulation of photons increases. In parallel, on the (right), a two-dimensional Airy wave packet observed, forming a spatiotemporal Airy photon.}
\label{Fig1}
\end{figure*}

\textbf{Experimental setup and general scheme.} An elongated cloud of Doppler cooled rubidium atoms, as shown in Fig.~\ref{Fig1}a, is prepared to the lowest hyperfine manifold $\vert 1 \rangle$ in a dark-line two-dimensional magneto-optical trap (MOT). After MOT loading, the atoms are released and a \SI{1.0}{\milli\second} of biphoton generation via spontaneous four-wave mixing (SFWM) takes place.  A pump, $\omega_p$, and a control beam, $\omega_c$, generate entangled Stokes, $\omega_S$, and anti-Stokes photons, $\omega_{AS}$, at wavelengths of \SI{780}{\nano\meter} and \SI{795}{\nano\meter} respectively, as shown in the four-level system of Fig.~\ref{Fig1}b. The circularly polarized coupling and pump beams are counter-propagating, with their axis being at a \SI{3.0}{\degree} angle with respect to the longitudinal axis of the MOT, i.e., the collection axis of the biphotons. The transverse spatial profile of the pump beam is shaped into a one-dimensional Airy beam by Fourier transforming a cubic phase imprinted to light using a spatial light modulator (SLM), as shown in Fig.~\ref{Fig1}a. The Stokes photons, after spectral filtering, are detected by the time resolving silicon avalanche single-photon counting modules (SPCM), and their arrival acts as a trigger to herald the slow anti-Stokes photons. The heralded anti-Stokes photons are recorded in a time-resolved manner, with $\tau$ representing their relative temporal delay with respect to their partner Stokes photons. After the anti-Stokes photons are collected by a single-mode optical fiber, in a time scale much shorter than the biphoton coherence time, their spatial shape is manipulated by an SLM.

In our experimental setup, with a far-detuned and weak pump beam exciting the MOT, a pair of Stokes and anti-Stokes photons are simultaneously emitted via SFWM from single atoms located within the elongated cloud. The coupling beam has approximately a longitudinally uniform profile over the atomic cloud. Since the Stokes photons are far-detuned from the atomic resonance, they experience practically zero linear susceptibility and fly throughout the cloud with the speed of light. Conversely, the anti-Stokes photons are on-resonance, and experience the slowing effect stemming from electromagnetically induced transparency (EIT). If the EIT window is much narrower than the third-order nonlinear susceptibility spectrum, $\chi^{(3)}(\omega)$, then the biphoton wavefunction $\psi\left(\tau\right)$ is described by~\cite{Zhao2015a,Zhao2016a}:

\begin{equation}\label{fp}
    \psi\left(\tau\right) \propto \chi^{\left( 3\right)}\left(0\right) E_c E_p V_g  f_p\left(\frac{L}{2}-V_g\tau\right),
\end{equation}

\noindent where $V_g$, is the group velocity of the anti-Stokes photons that leads to a group delay $\tau_g = L/V_g={2 \gamma_{13} OD}/{\vert \Omega_{c} \vert ^{2}}$ in an L-long cold atomic ensemble with a natural linewidth $\gamma_{13}$ for the relevant transition $\vert 1\rangle \rightarrow \vert 3\rangle$. In this case the optical depth is $OD$, and $\Omega_c$ is the Rabi frequency of the coupling beam. Meanwhile $E_p$ and $E_c$ represent the electric field amplitudes for the pump and the coupling beam while $f_p\left(z\right)$ expresses the z-dependence of the electric field of the pump, i.e., $\Omega_p\left(z\right)=\Omega_p f_p\left(z\right)$, with $f_p\left(z\right)$ being normalized as $\int_{-L/2}^{L/2} \vert f_p\left(z\right) \vert^2 dz/L=1$.

Equation~\ref{fp} elucidates the physical mechanism determining the space-to-time mapping of the pump light into the desired waveform for the Stokes/anti-Stokes biphotons~\cite{Zhao2015a}. We note that the Stokes photons traverse the atomic medium at the vacuum speed of light, while the velocity of the heralded slow anti-Stokes photons depends predominantly on the density of the optically thick cold atomic ensemble and the intensity of the coupling beam. As a result, it is the relative time delay in the arrival of the slow anti-Stokes photons in each photon pair that determines the temporal biphoton wavefunction. The space-dependent pump beam controls the generation rate of photon pairs and as a result the temporal biphoton waveform. Alternatively, temporal amplitude and multiple-$\pi$-phase modulation could be applied to the emitted photons but with the cost of extra photon losses~\cite{EOM2008}. When manipulating the generation rate along the elongated atomic cloud, we effectively apply a spatial multiplexing scheme. In other words, by utilizing the slow light effect, the spatial information is transformed into a temporal one for the photon pairs.

Shaping the spatial wavefunction of the generated Airy biphotons can conveniently rely on methods from linear classical optics, and hence, the (2+1)D biphoton source reported here can be exploited for generating a number of nonclassical spatiotemporal wave packets, such as X-waves~\cite{DiTrapani2012}, abruptly autofocusing wave packets~\cite{Papazoglou2011}, and even, nonspreading Helmholtz photons~\cite{Kaminer2012,Aleahmad2012}. 

The three-dimensional space-time Airy wave packet considered here has the separable form:

\begin{equation}
    \vert\Psi_{xy\tau}\rangle_{z=0} = \prod_{s = x,y,\tau}{Ai\left(s/s_0\right)\exp\left( a_s s\right)},
\end{equation}
where $s_0 = x_0, y_0,$ and $\tau_0$ represent the spatial and temporal widths of the single-photon Airy wavefunction, while the $a_s$ for $s=x,y,\tau$ are the corresponding small ($a\ll1$) and positive exponential decay parameters. This nonclassical photon wavefunction inherits the remarkable properties of the Airy wave packet~\cite{Siviloglou2007a, Efremidis2019}, such as nondiffraction and transversal self-bending for the spatial mode. Importantly, it also exhibits nondispersion and longitudinal acceleration for the temporal component, and self-healing as was demonstrated in~\cite{Broky2008}, and only recently in the quantum realm~\cite{Li2021}.

Figure~\ref{Fig1}c conceptually illustrates the formation of nonclassical spatiotemporal Airy wave packets, photon-by-photon. After the detection of every trigger photon, the wavefunction simultaneously collapses to a single event in the spatial and the temporal domain. As expected, the accumulation of photons leads to a smooth Airy pattern. 

\begin{figure*}[htbp]
  \centering
  \includegraphics[width=1.0\linewidth]{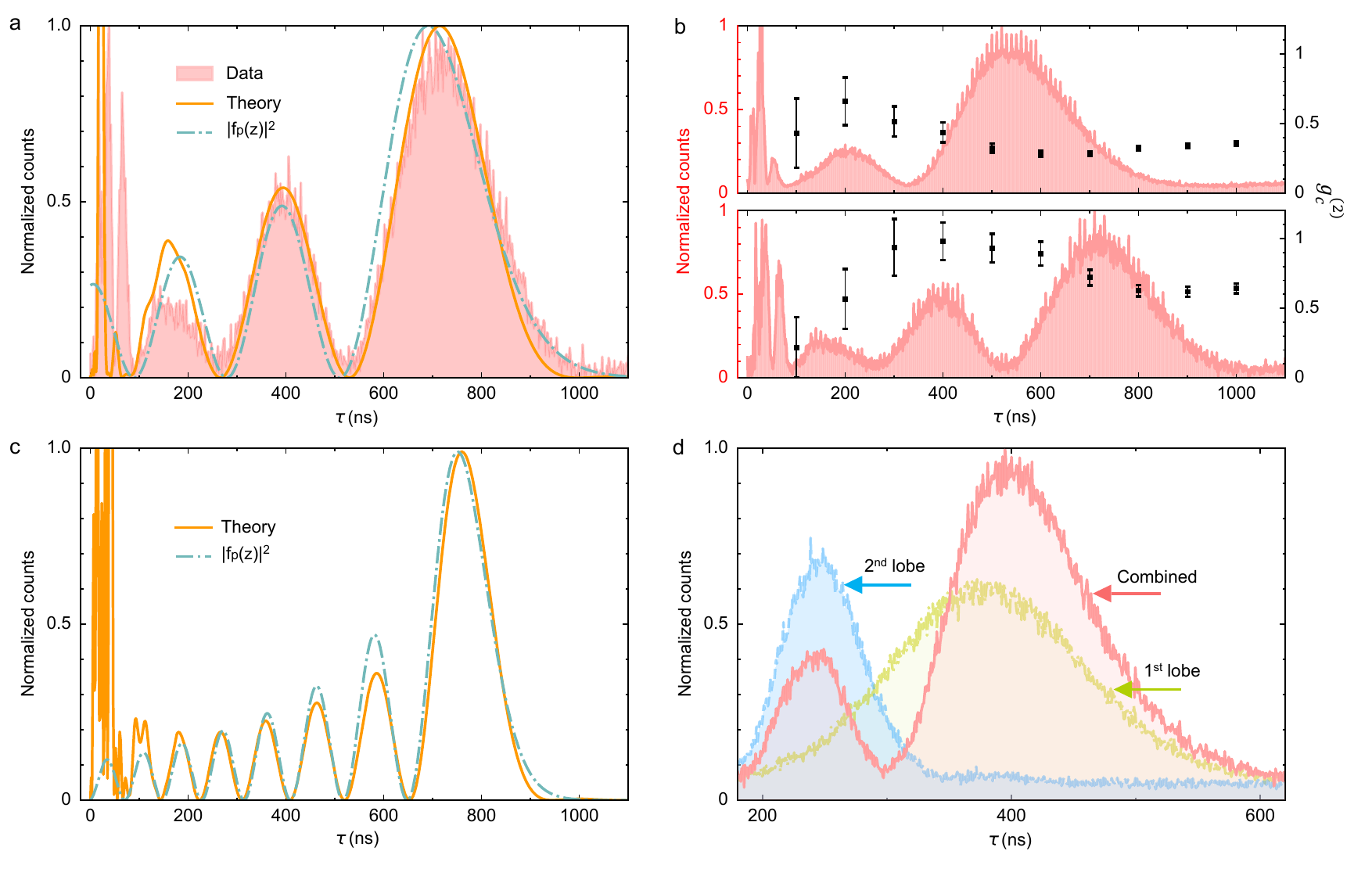}
 \caption{$\mathbf{\vert}$\textbf{ Observation of temporal Airy biphotons.} \textbf{a,} Temporal Airy biphoton waveform. Normalized counts of the observed biphotons (shaded pink), numerically predicted waveforms (orange solid line) based on the interaction picture for $OD = 150$ and $\Omega_c =2\pi \times \SI{10}{\MHz}$, and analytical prediction (dash dotted green line) from the Airy pump shape $f_p\left(z\right) = Ai\left(z/z_0\right)\exp\left(a z/z_0\right)$ with $z_0 = \SI{18.7}{\milli\meter}$, and $a \approx 0.1$. \textbf{b,} (top) Two- and (bottom) three-lobe Airy wave packets (shaded pink) and their corresponding self-correlations, $g_c^{\left(2\right)}\left(\tau\right)$ (black squares), from the Hanbury Brown-Twiss experiment. Nonclassical light has $g_c^{\left(2\right)}\left(\tau\right)<1$. The error bars are calculated from the one standard deviation of the photon counts (see Supplementary Material). \textbf{c,} Simulated (dash dotted green line), and analytical (orange solid line) Airy waveforms for high optical depths ($OD = 500$). \textbf{d,}  Locally probing the biphoton wavefunction phase. Observed normalized counts when the Airy pump is imposed as a whole (shaded pink), with only the first lobe (shaded green), and only the second lobe (shaded blue). The alternating phase leads to zero counts approximately at the intersection of the first and second lobe waveforms. For normalization purposes we set the maximum of the group delayed waveform to 1.}
\label{Fig2}
\end{figure*}

\begin{figure*}[htbp]
\centering
\includegraphics[width=0.8\linewidth]{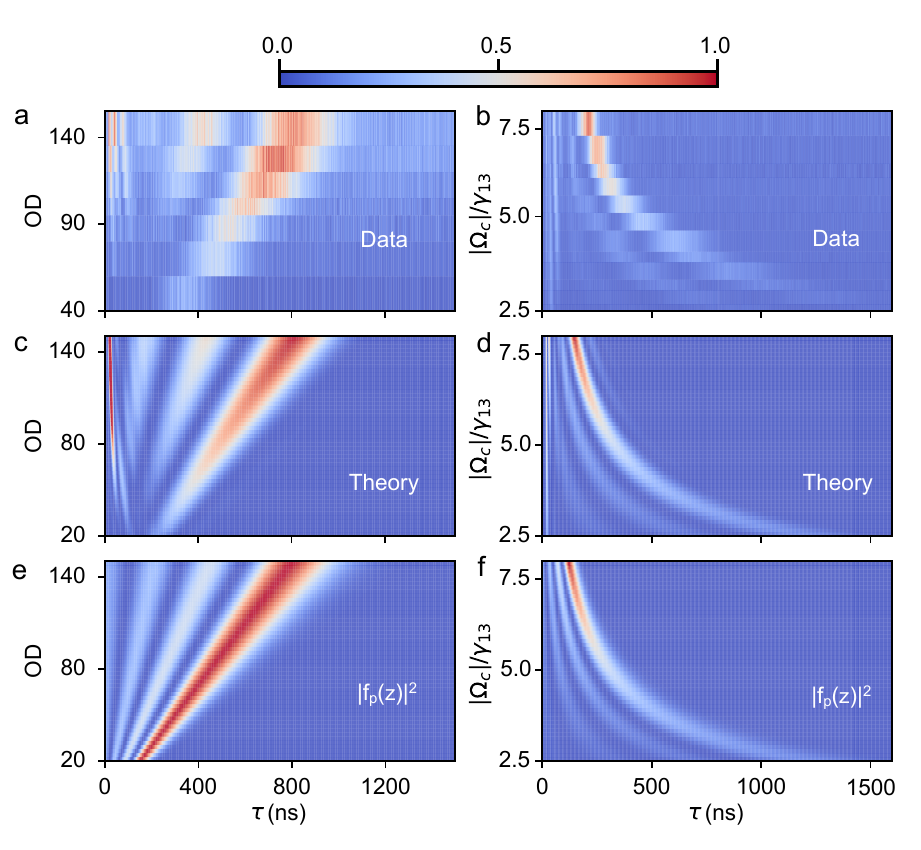}\caption{$\mathbf{\vert}$\textbf{ Engineering the temporal quantum Airy wavefunction $\psi\left(\tau\right)$.} \textbf{a, c, e,} Temporal waveform density $\vert \psi\left(\tau\right)\vert ^2$ as a function of optical depth $OD$. The Rabi frequency of the coupling beam is fixed to $\Omega_c = 2\pi\times\SI{10}{\MHz}$. \textbf{a,} Experiment, \textbf{c,} simulation, and \textbf{e,} analytical prediction from $f_p\left(z\right)$. \textbf{b, d, f,} Temporal waveform as a function of the normalized coupling Rabi frequency $\vert \Omega_{c} \vert/\gamma_{13}$. The optical depth of the atomic cloud is set to $OD = 150$. \textbf{b,} Experiment, \textbf{d,} simulation, and \textbf{f,} analytical prediction from Eqn.~\ref{fp}. These results are consistent with the theoretical calculation based on the photon delay $\tau_g = {2 \gamma_{13} OD}/{\vert \Omega_{c} \vert ^{2}}$. For short time delays, precursor photons appear as shown in our experimental (panels a and b) and theoretical results (panels c and d).}
\label{Fig3}
\end{figure*}

\begin{figure*}[htbp]
\centering
\includegraphics[width=1\linewidth]{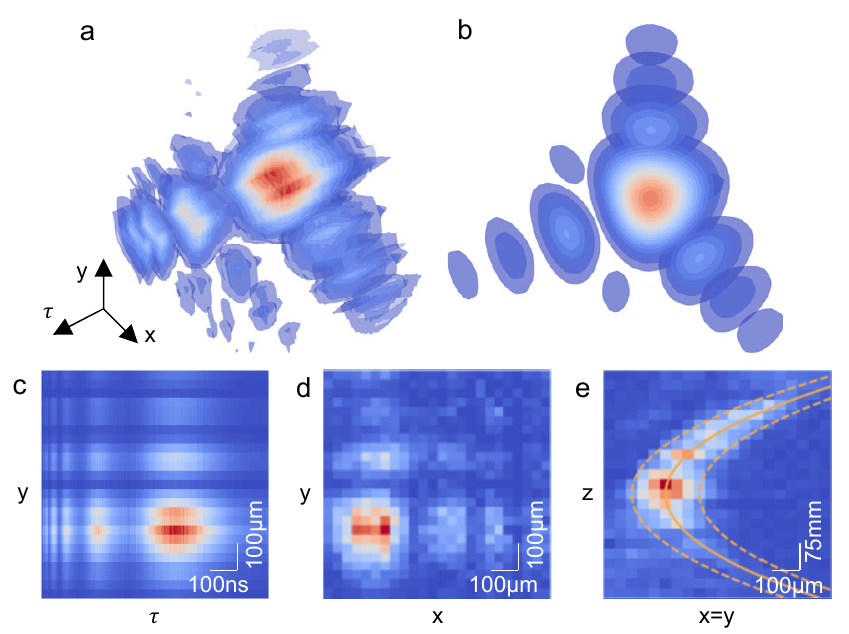}
\caption{$\mathbf{\vert}$\textbf{ A three-dimensional spatiotemporal Airy photon.} Isosurface of \textbf{a,} the experimentally observed squared amplitude $\vert \psi_{xy\tau} \vert ^2$ of the Airy biphotons, and \textbf{b,} the corresponding theoretical spatiotemporal wave packets. $17$ isosurfaces ranging from values $0.1$ to $1.0$ are plotted. The vector lengths represent $x,y$, and $\tau$ scales of $\SI{100}{\micro\meter}$, $\SI{100}{\micro\meter}$, and $\SI{200}{\nano\second}$ respectively. \textbf{c,} Observed photon shape (normalized counts) along the $x-$axis. A projection of the temporal $\tau$ and the spatial $y$ photon probability density along the $x-$axis. \textbf{d,} Observed photon shape along the $\tau-$axis. A projection of the temporal two-dimensional spatial photon probability density along the time axis $\tau$. \textbf{e,} Spatial self-bending of spatiotemporal single Airy photons. The $z-$propagation of the nonclassical Airy photons is observed along the diagonal ($x=y$) cross-section. The theoretically expected parabolic trajectory (solid for the maximum and dashed for the full width half-maximum) is overlaid on the recorded data.}
\label{Fig4}
\end{figure*}

\begin{figure*}[htbp]
\centering
\includegraphics[width=1.0\linewidth]{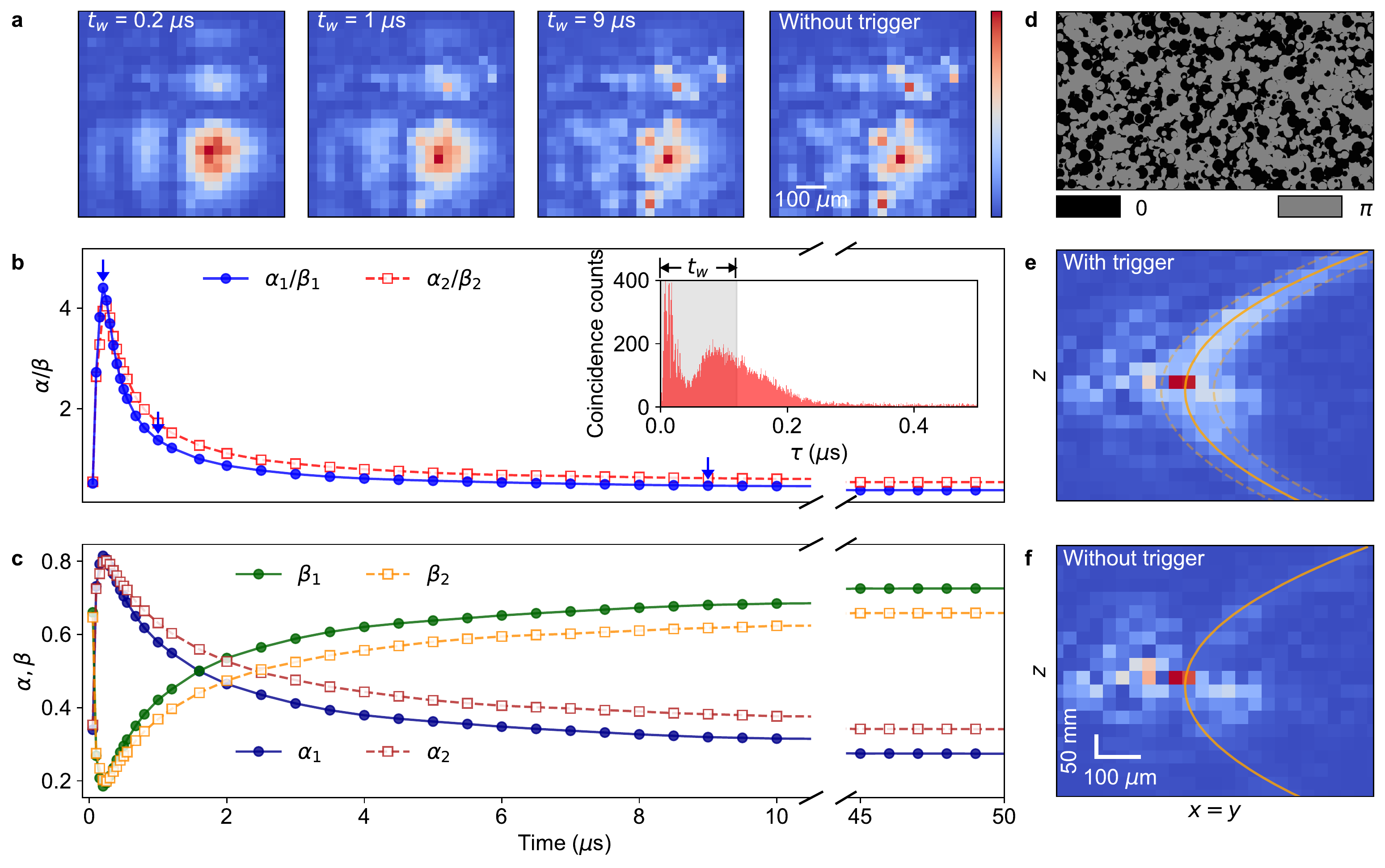}
\caption{$\mathbf{\vert}$\textbf{ Concealing correlated nondiffracting photons in classical light noise.}   \textbf{a,} Spatial Airy pattern emerging from noise. As the time window $t_w$ of collecting heralded photons is increased (here from $\SI{0.2}{\micro\second}$ to $\SI{1}{\micro\second}$ and then $\SI{9.0}{\micro\second}$) the characteristic Airy pattern is concealed by noise. For untriggered collection (right) noise dominates the image. \textbf{b,} Airy to noise ratio $\alpha / \beta$ as a function of $t_w$. The blue arrows correspond to the first three images of (a). Blue and red lines show two measurements with different classical noise powers. The inset depicts a typical biphoton waveform for a Gaussian pump. \textbf{c,} Normalized projections of Airy photons and classical noise. Solid dots and empty squares correspond to the measurements of (b). For time windows narrower than approximately $\SI{2}{\micro\second}$ the Airy photon pattern emerges. \textbf{d,} Phase mask for classical speckle generation. A binary phase with only $0$ and $\pi$ randomly distributed disks is imprinted on a classical Gaussian beam to generate via Fourier transformation the noise pattern. \textbf{e,} Propagation of nondiffracting correlated Airy photons over classical noise. When triggering of Stokes photons is used the self-bending of Airy photons is clearly visible. A contour theoretical prediction for the peak and the full width half maximum curves is overlaid. \textbf{f,} Propagation of uncorrelated thermal Airy photons over classical noise. When no triggering is employed the strong noise masks the Airy light completely. The solid line indicates the position of the maximum of the first Airy lobe. The theoretically expected parabolic trajectory (orange dashed line) is overlaid on the recorded data. For both \textbf{e} and \textbf{f} an $x=y$ cut of the collected light (Airy photons+noise) was measured.}

\label{Fig5}
\end{figure*}

\noindent
\textbf{Mapping a 1D Airy beam onto the photon temporal waveform.} To generate temporal Airy biphotons, we use a one-dimensional classical Airy-shaped pump (Fig.~\ref{Fig2}a). The characteristic oscillating and asymmetric Airy shape is clearly observed, in excellent agreement with the numerical simulations carried out within the interaction picture (see Methods). In addition our results conform well with analytical predictions based on the space-to-time mapping of the longitudinal profile $f_p\left(z\right)$ of the pumping beam for an average group velocity of $V_g\approx{0.000076c}$ in a medium with optical depth of $150$,  and a peak Rabi frequency for the coupling beam of $\Omega_c = 2\pi \times \SI{10}{\MHz}$. We note that, in the first \SI{30}{\nano\second} after the Stokes trigger, a precursor~\cite{Chen2010a} pulse that is not temporally modulated by the Airy pump beam and travels at the vacuum speed of light is observed. 

To further probe the nonclassical features of the generated Airy photons, we pass them through a fiber beam splitter to perform Hanbury Brown-Twiss interference and their self-correlation for two different pump shapes was measured as a function of the integration window $g^{\left(2\right)}_c$, as shown in Fig.~\ref{Fig2}b. In both cases we measure $g^{\left(2\right)}_c<1$, thus verifying their nonclassical nature.  We note that the self-correlation of the generated photons could, in principle, be quadratically improved by reducing the pump power, but at a  cost of lower photon rates. The atomic medium can support even denser time-binning compared to the one demonstrated here if the atomic density was further increased. Ultra-high optical depths of up to a few thousands have been demonstrated~\cite{Sparkes2013} and in the case of an ensemble with $OD=500$, a temporal Airy biphoton, as the one displayed in Fig.~\ref{Fig2}c, would be observable. Furthermore, to verify the spatial origin of the temporal lobes and provide additional evidence for the manipulation of the amplitude/phase of the complex cross-correlation $g^{\left(2\right)}_c$ corresponding to the generated biphotons, we have performed additional experiments as shown in Fig.~\ref{Fig2}d. The temporal profile of the Airy wavefunction is observed in the case where the main or the secondary lobes are blocked. The interference of single photons with a $\pi$ phase difference leads to zeroing of the wavefunction amplitude, approximately at the location where the first and the second lobe would be equal in isolation. Additional simulations for a pumping beam having an Airy intensity profile but a flat phase, are presented in the Supplementary Material.

To establish the robustness and versatility of the implemented method for generating nonclassical temporal Airy wave packets, we have engineered the biphoton shape when the two most relevant parameters for the photon generation and photon delay can vary, i.e., the optical depth, $OD$, of the atomic medium and the coupling laser Rabi frequency, $\Omega_c$. The experimental results for the Airy biphoton wavefunction for different optical depths and coupling powers are presented in Figs.~\ref{Fig3}a,b, while the corresponding numerical simulations, and theoretical predictions based on Eqn.~\ref{fp} are shown for the same range of scanned parameters in Figs.~\ref{Fig3}c,d and Figs.~\ref{Fig3}e,f respectively. The demonstrated quantitative agreement over such a large range of experimental parameter space clearly indicates that the pulse shape and the photon rate can be tuned precisely in order to either stretch or compress the single-photon temporal wavefunctions~\cite{Sedziak2019, Li2015}.

\noindent
\textbf{Observation of spatiotemporal Airy photons.} To spatially control the generated biphotons, we collect the Stokes photons with a single-mode fiber, while the heralded photons pass through another fiber and are shaped in two transverse spatial dimensions. After exiting the single-mode fiber, the anti-Stokes photons are spatially magnified three times to a beam with a waist of \SI{5}{\milli\meter}. Consequently, a two-dimensional cubic phase is imprinted on the wavefront using an SLM that allows one in turn to shape it by means of standard Fourier transformation into two-dimensional Airy photons. We note here that the photons arrive one-by-one and the Airy pattern is formed clearly only after several thousands of photons events are registered. To image the quantum Airy beam, an additional linear phase is added in the SLM phase pattern whose dual role is to separate the undiffracted zero order of unmodulated photons, and most importantly, to provide a precise way to scan along the two transverse dimensions the probability density of the Airy photons (see Supplementary Material).

In Figs.~\ref{Fig4}a,b the experimentally obtained three-dimensional isosurfaces for the spatiotemporal Airy photons are presented along with the corresponding ideal theoretical wave packets for comparison. We note that this is the first realization of a single-photon nonspreading optical bullet in the quantum realm. This kind of (2+1)D Airy wave packets have been observed in the classical regime of extreme nonlinear optics where the photons per pulse can exceed $10^{14}$~\cite{Tzortzakis2010}. 

The projection of the spatiotemporal Airy photons on the temporal, $\tau$ and spatial dimension $y$, as seen from the $x-$direction, is shown in Fig.~\ref{Fig4}c, while the complementary spatial photon shape is shown in Fig.~\ref{Fig4}a where $a=0.1$ and $x_0=\SI{100}{\micro\meter}$. To increase the rate of the collected photons we have increased the exponential containment of the Airy wave packets. This does not lead to a noticeable distortion of the classical self-bending of the Airy wave packets, as shown in Fig.~\ref{Fig4}e. The Fresnel propagation of the single-photons is observed by introducing an additional quadratic phase, and their spatial wavefunction is scanned along the diagonal $x=y$ (see Supplementary Material).

For observing spatiotemporal dynamics as in~\cite{Chong2010}, dispersion and diffraction lengths should be comparable. This can be achieved by either increasing the size of the transverse spatial features, or by decreasing the temporal pulse widths through an increase in the coupling power. These properties can also be revealed in atomic media with steep spectral features or alternatively by using other methods like spontaneous down-conversion that relies on faster optical nonlinearities.  

\noindent
\textbf{Concealing single Airy photons in light noise.} As shown above, the combination of space-to-time mapping in a slow-light medium, and the simultaneous independent control of the photon beam shape enables us to freely synthesize experimentally nonclassical spatiotemporal wave packets. The nondiffracting and accelerating properties of the spatial Airy beams can now be complemented and enhanced by their heralded quantum realization. We have taken advantage of photon correlations to camouflage the Airy-shaped photons in a speckled background of classical photon noise. This coherent noise source is derived from a highly attenuated laser with the same frequency as the Airy single photons and is characterized by intensity features comparable in size to the lobes of the Airy beam, while its average intensity evidently exceeds it. In the first three panels of Fig.~\ref{Fig5}a it is clear that an observer with access to the Stokes trigger, which heralds the Airy photons, can retrieve the single-photon spatial Airy wavefunction at low integration times with high fidelity, while an observer without access to these trigger photons will detect a severely distorted pattern that is dominated by the noise intensity features, as indicated in the last panel of Fig.~\ref{Fig5}a. Fig. ~\ref{Fig5}b quantitatively demonstrates this crossover from the quantum to the classical regime where the noise dominates the detected signal. $\alpha/\beta$ represents the ratio of photons in the Airy, $\alpha$, and in the classical noise pattern, $\beta$, as a function of the integration window. The arrows mark the corresponding panels in Fig. ~\ref{Fig5}a. We note that an observer without any access to the trigger photons is forced to detect photons at uncorrelated time intervals, which results in noise-dominated patterns as represented by the horizontal asymptotes in Figs.~\ref{Fig5}b,c. To prove the robustness of this camouflaging scheme we have repeated this experiment for different noise levels and observed similar curves. It must be noted that this behavior is not unique to nondiffracting beams. The novelty here is that photon correlations can enhance the resilience of nonspreading propagation dynamics of the spatial Airy photons, even in the presence of a strong localized noise source, created using the binary phase of Fig.~\ref{Fig5}d. The resilience of Airy photons to diffraction allowed us to increase even further (a factor of $5$) the ratio of classical noise to Airy photons, and the propagation dynamics of the Airy photons in the noise photon background is presented at short and at very long times in Figs.~\ref{Fig5}e,f. It is evident that the nondiffracting Airy photons can be revealed after propagation as shown in Fig.~\ref{Fig5}e. In Fig.~\ref{Fig5}f it is clear that the noise is so strong that for an observer who does not have access to the Stokes photons the detected pattern will appear completely like noise.

\section{Discussion and conclusions}\label{sec12}
In this work, we have demonstrated for the first time spatiotemporal control of single quanta in the form of nonspreading quantum Airy photons with long coherence times, by merging nonlinear quantum optics in cold atomic ensembles and nondiffracting spatial photonics. A mapping of the longitudinal modulation of a classical pumping beam has enabled us to sculpt the complex temporal wave-function of the heralded photons in a slow-light medium, while the spatial degrees of freedom were controlled by subsequent transverse shaping of the generated single photons. So far, these two directions of quantum optics based on nondiffracting beams and ensembles of cold neutral atoms have not been simultaneously pursued. As to future directions, of interest will be to use a second atomic ensemble or a crystal~\cite{Wang2019} with steep spectral features, as the ones encountered in electromagnetically induced transparency, to observe the spatiotemporal dynamics of single-photon bullets. Based on our scheme, shaping and propagation invariance of temporal quantum Airy waveforms in the femtosecond regime can be demonstrated by utilizing ultrafast laser techniques~\cite{Weiner2000, Peer2005, Chong2010}. Spatiotemporal self-healing of entanglement, with applications in quantum networks can be observed as demonstrated in the spatial domain~\cite{ McLaren2014b, Wang2022}. Finally, generating the first nonparaxial single-photon bullets is also another exciting possibility~\cite{Kaminer2012, Aleahmad2012}. Our work can shed light on important quests such as the resilience of entanglement in atomic media under the action of self-healing, and can lead to advances in imaging, and multimode information storage and encoding in the quantum regime~\cite{PhysRevLett.129.193601, Hang2014}.

\clearpage
\clearpage
\section*{Methods}\label{sec11}

\noindent
\textbf{Cold atoms experimental setup.} 
An elongated \SI{1.8}{\centi\meter} atomic cloud of $^{85}$Rb with an aspect ratio of 36:1 is laser cooled to \SI{120}{\micro\kelvin} in a two-dimensional dark-line magneto-optical trap (MOT). 
Its optical depth ($OD$) can reach $250$ and continuously tuned by the power of the repumping light on the $ \vert 5S_{1/2},F=2 \rangle \rightarrow \vert 5P_{3/2},F'=2\rangle$ transition. 
The repetition rate of the experiment is \SI{100}{\Hz}. 
A \SI{9}{\milli\second} loading stage brings all the atoms to the ground state hyperfine manifold: $\vert1\rangle=\vert 5S_{1/2},F=2 \rangle$ and is followed by \SI{1.0}{\milli\second} of biphoton generation at \SI{780}{\nano\meter} and \SI{795}{\nano\meter} for the Stokes ($\vert4\rangle=\vert 5P_{3/2},F'=3 \rangle \rightarrow \vert2\rangle = \vert5S_{1/2},F=3\rangle$ ) and the anti-Stokes ($\vert3\rangle = \vert 5P_{1/2},F'=3\rangle \rightarrow \vert1\rangle$) photons respectively. 
The classical coupling beam has a waist of $\SI{1.2}{\milli\meter}$, and it is on resonance with the $\vert2\rangle$ to $\vert3\rangle$ transition, 
while the counter-propagating Airy-shaped pumping beam is blue-detuned by $\Delta_{p}=\SI{120}{\MHz}$ from the $\vert1\rangle$ to $\vert4\rangle$ transition.
The Rabi frequencies $\Omega_{c}$ and $\Omega_{p}$ are kept to $3.3\gamma_{13}$ and $0.3\gamma_{13}$ throughout the experiments, unless otherwise stated. The relevant dephasing rates are $\gamma_{13}=2\pi\times\SI{3}{MHz}$, and  $\gamma_{12}=2\pi\times\SI{30}{kHz}$. The latter is associated with the dipole-forbidden transition $\vert 1 \rangle \rightarrow \vert 2 \rangle$.

\noindent
\textbf{Simulation of the temporal biphotons.} 
The temporal wavefunction $\psi \left(\tau\right)$ is simulated by: 
\begin{equation}
\psi\left( \tau\right) = \frac{\sqrt{\omega_{s0}\omega_{as0}}}{i 4\pi c}\int{F\left(\omega\right) Q\left(\omega \right) e^{-i\omega \tau} d\omega},
\end{equation}
with

\begin{equation}
F\left(\omega\right) \equiv \int_{-L/2}^{+L/2}{ \chi^{\left( 3\right)}E_pE_ce^{-i\int_{0}^{z}{\Delta k \left(z'\right)dz'}}dz}
,    
\end{equation}
representing the biphoton generation due to the third-order interaction in the nonlinear atomic medium, and the overall wavevector mismatch while
\begin{equation}
Q\left(\omega\right) \equiv e^{i\int_{0}^{L/2}\left[k_s\left(-z'\right)+k_{as}\left(z'\right)\right]dz'}
,
\end{equation}
expresses the accumulating phase on the two counter-propagating generated photons.
We note that, in general, the third-order nonlinearity and the classical beams are z-dependent. 

\noindent
\textbf{Spatial shaping of the single photons.} 
The fiber-collected anti-Stokes photons are expanded to a Gaussian beam with a waist of $w_0=\SI{5}{\milli\meter}$, and subsequently are Fourier-transformed to spatial Airy wave packets by the combination of an $f=\SI{400}{\milli\meter}$ lens, and a high diffraction efficiency ($92\%$) spatial light modulator (SLM, Holoeye PLUTO-2). The SLM, with a pixel size $d = \SI{8.0}{\micro\meter}$, imprints on the photons a two-dimensional phase of the third-order polynomial form: $\sum_{n=1,2,3}^{}{\left(\Delta_{n,x'}x'^n+\Delta_{n,y'}y'^n\right)}$, where the transverse coordinates on the SLM plane are $x'=i\cdot d$ and $y'=j\cdot d$, and $i,j$ are the corresponding pixel indices. For simplicity, we skip the coordinate indices and give the expressions only for the $x'$. The linear phase term separates the Airy modulated photons from the undiffracted ones, and shifts them transversely by ${\Delta_{1} \lambda f}/{2\pi}$. 

The quadratic phase leads to a Fresnel-equivalent propagation distance  ${\lambda f^2 \Delta_2}/\pi$. 

The cubic phase generates at the origin $z=0$ the spatial Airy pattern $Ai\left(x/x_0\right)\exp\left(\alpha_x x\right)$ with $x_0={(3\Delta_3)^{1/3} \lambda f}/{2\pi}$, and $\alpha_x=1/\omega_0^2 (3\Delta_3)^{2/3}$. 

\noindent
\textbf{Generating the speckled light pattern.} The speckled light noise is generated by Fourier transforming an attenuated laser beam modulated by a random binary ($0/\pi$) phase pattern from a high resolution SLM (Holoeye, GAEA-2). We use a distribution of $4000$ disks randomly positioned in the whole area of the SLM with radii in the range of $\SI{160}{\micro\meter}$ to $\SI{640}{\micro\meter}$. To avoid the influence of the zero-order beam diffracted by the SLM, we additionally apply blazed grating which is not shown in Fig.~\ref{Fig5}d. All the scanning shifts for the noise pattern are the same as for the Airy photons.  

\noindent
\textbf{Signal-to-noise ratio for the biphotons.}
For a Stokes trigger the count distribution of the heralded anti-Stokes photons is peaked over the first hundreds of nanoseconds, while the noise photons arrive at a constant rate and therefore the quantum correlations lead to a significant enhancement of the Airy signal-to-noise ratio $a(w_t)/b(w_t)$. The evolution of the Airy and noise signals is quantified by extracting the corresponding projection coefficients $\alpha\left(t_w\right)$ and $\beta\left(t_w\right)$ from the detected normalized spatial distribution of all the heralded photons within a time window $t_w$: $N_{total}\left(x,y, t_w\right) = \alpha\left(t_w\right)N_{Airy}\left(x,y\right)+\beta\left(t_w\right)N_{noise}\left(x,y\right)$. For unity normalization of the basis functions $N_{Airy}\left(x,y\right)$ and $N_{noise}\left(x,y\right)$, and since all the other stray photon sources were negligible $\alpha+\beta = 1$. 

\backmatter

\section*{Acknowledgments}
This work is supported by the National Natural Science Foundation of China (NSFC) through Grants No. 12074171, No. 12074168, No. 92265109 and No.12204227; the Guangdong Provincial Key Laboratory (Grant No. 2019B121203002);
the Guangdong projects under Grant No.2022B1515020096 and No. 2019ZT08X324. The work of DNC was partially supported by ONR MURI (N00014-20-1-2789), National Science Foundation (NSF) (DMR-1420620, EECS-1711230), MPS Simons collaboration (Simons grant 733682), W. M. Keck Foundation, US–Israel Binational Science Foundation (BSF: 2016381), and the Qatar National Research Fund (grant NPRP13S0121-200126).

\bigskip

\clearpage

\nolinenumbers
\bibliography{sn-bibliography}



\section*{Supplementary material}

# Supplementary material for "Spatiotemporal single-photon Airy bullets"

## S.1 Cold atomic ensemble preparation

To trap and laser cool the $^{85}$Rb atoms three orthogonal pairs of counter-propagating circularly polarized trapping beams with a total power of 120 mW and a detuning of $-20$ MHz from the cycling transition $|5^2 S_{1/2}, F = 3\rangle \rightarrow |5^2 P_{3/2}, F' = 4\rangle$ are combined with two orthogonal repumping beams of 20 mW resonant with the $|5^2 S_{1/2}, F = 2\rangle \rightarrow |5^2 P_{3/2}, F' = 2\rangle$ transition. All these beams have a Gaussian shape with a waist of 12 mm and they intersect on the zero-field line parallel to the long axis of a rectangular coil that creates a magnetic field gradient of 7 G/cm. To form the dark-line magneto-optical trap (MOT) that leads to an enhanced optical depth ($OD$) compared to the standard two-dimensional MOT the central lines of the repumping beams are blocked. Cigar-shaped atomic ensembles with typical temperatures of 120 μK, an approximate length of 1.8 cm, and optical depth that can reach 250 are obtained.

## S.2 Time sequence for the biphoton generation

For the efficient generation and detection of correlated pairs of Stokes and anti-Stokes photons with long coherence time, the MOT light has to be switched off. The measurements are performed in a pulsed mode with a repetition rate of 100 Hz as shown in the time sequence of Fig. 1a. During the 9.0 ms of MOT loading the trapping beams are kept on while the repumping beams are turned off 0.3 ms before the trapping ones to prepare all the atoms to the $F = 2$ ground state manifold. After a few microseconds the biphoton generation or the atomic density characterization take place, depending on the experiment that we are performing. The 795 nm coupling and the 780 nm pumping beams enter the atomic cloud to generate via spontaneous four-wave mixing (SFWM) the entangled photon pairs as illustrated schematically in Figs. 1a and b of the main text. In Figs. 1b and c typical waveforms of the generated biphotons are shown when pump beam has Gaussian or Airy shape respectively.

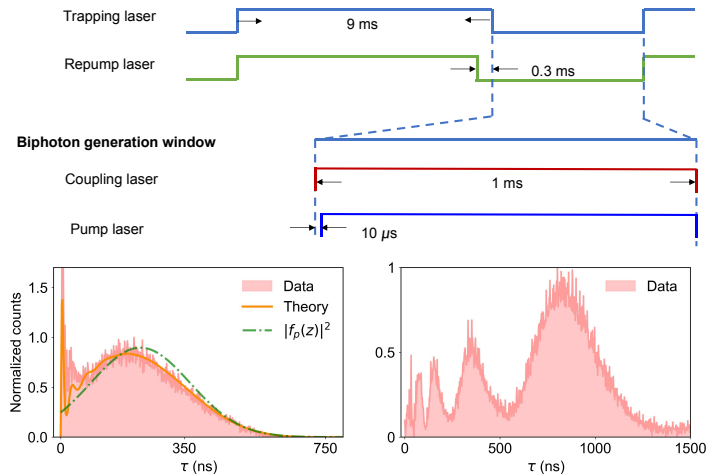

**Fig. 1 | Biphoton generation time sequence. a,** The time sequence consisting of 9.0 ms of cold atomic ensemble preparation, and 1.0 ms of biphoton generation by SFWM. Typical temporal biphoton waveforms for a pump beam **b,** with a Gaussian longitudinal profile $f_p(z)$, and **c,** a one-dimensional Airy shape.

## S.3 Time sequence for the atomic density measurement

To characterize the optical depth of the atomic cloud, and the dephasing rate $\gamma_{12}$, which is associated with the coherence time of the generated biphotons, we use electromagnetically induced transparency (EIT) with the time sequence of Fig. 2a. The three-level EIT scheme with the coupling and the probe laser beams (same frequency with the anti-Stokes photons) is shown in Fig. 2b. The probe ($w_0 = 125\,\mu m$) and the control beam are pulsed for 100 μs. To obtain the characteristic EIT spectra the frequency of the weak (power less than 20 nW) probe light is swept linearly from $-40$ MHz to $+40$ MHz with respect to the $|1\rangle \rightarrow |3\rangle$ transition. The typical EIT spectrum and the corresponding fit are shown in Fig. 2c. From the fitting results we extract the optical depth, the dephasing rates, and the Rabi frequency of the coupling beam. The optical depth can be controlled experimentally by reducing the power of the repumping light during the atom loading stage or by varying the current of the alkali dispensers.

## S.4 Photon detection and characterization.

The Stokes and anti-Stokes photons, after being emitted from the two opposite ends of the elongated atomic ensemble, are collected by antireflection coated single-mode polarization maintaining fibers. Their polarization is converted from circular to linear by the combination of a quarter waveplate and a polarizing beamsplitter. The collected photons are spectrally purified by a 1 nm bandpass filter and a custom-made low-finesse Fabry-Pérot cavity with a linewidth of 500 MHz before their detection by multimode fiber-coupled single-photon counting modules (SPCM, Excelitas SPCM-AQRH-14-FC). The main sources of stray photons are scattered pumping and coupling photons, as well

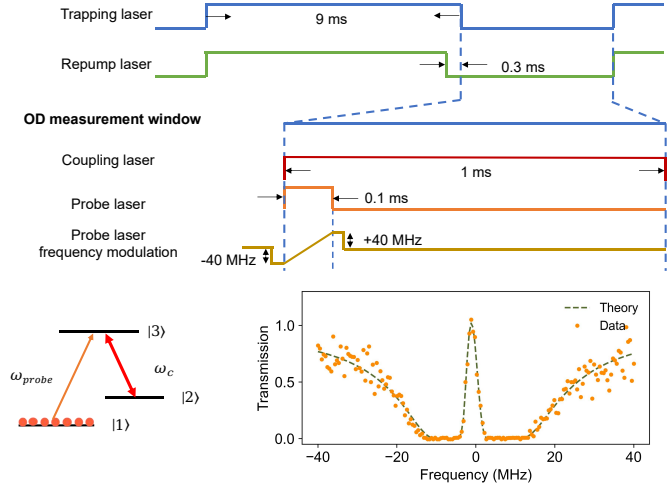

**Fig. 2 | Optical depth measurement time sequence. a,** The time sequence consisting of 9.0 ms of cold atomic ensemble preparation, and 0.1 ms of EIT frequency scanning. **b,** A schematic of the three-level configuration used in our experiment, and **c,** a typical experimental spectrum for an atomic cloud with $OD = 100$ retrieved from the associated fitting curves. $|1\rangle = |5^2S_{1/2}, F = 2\rangle$, $|2\rangle = |5^2S_{1/2}, F = 3\rangle$, and $|3\rangle = |5^2P_{1/2}, F' = 3\rangle$.

as thermal unpaired photons emitted from the cold atomic ensemble. The dark counts of the detectors are negligible throughout this work. A time resolution of 164 ps is in principle achievable according to the SPCM specifications. For our experiments, the typical Stokes and anti-Stokes photon rates can range from sub-kHz to a few kHz, depending on the shape and the power of the classical beams, as well as the atomic density among others.

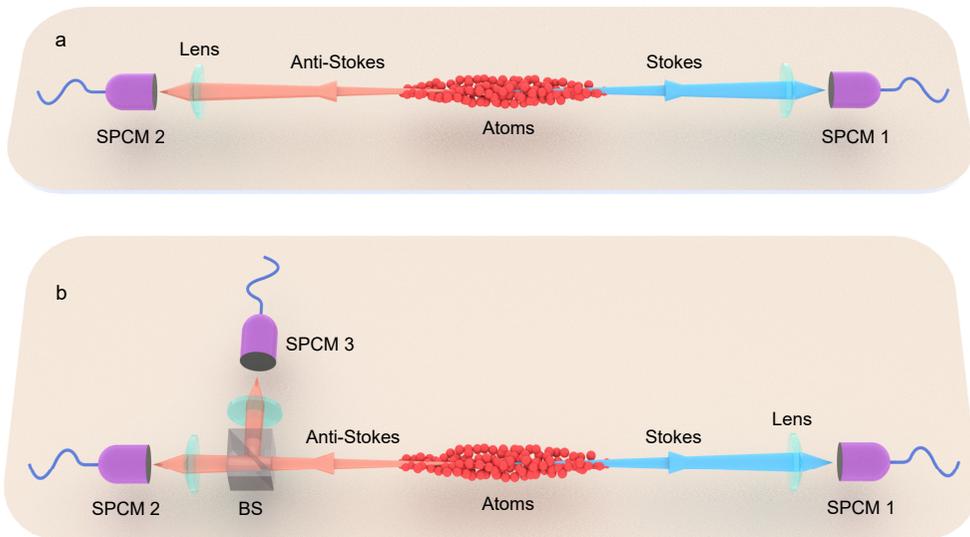

**Fig. 3 | Optical setups for the characterization of the Airy photons.** Optical configurations for measuring the **a,** temporal waveforms of the biphotons, and the **b,** self-correlation $g_c^{(2)}(\tau)$ of the emitted anti-Stokes photons.

To measure the biphoton waveforms reported in the main text one Stokes and one anti-Stokes SPCM are used to collect the corresponding trigger and heralded photons, as shown in Fig. 3a. The self-correlation $g_c^{(2)}(\tau)$ of the emitted anti-Stokes photons can be measured by the Hanbury Brown-Twiss setup of Fig. 3b. The self-correlation function $g_c^{(2)}(\tau)$ for the anti-Stokes photons is expressed as:

$$g_c^{(2)}(\tau) = \frac{N_1 N_{1,23}}{N_{1,2} N_{1,3}}, \tag{1}$$

where $N_1$ is given by the trigger Stokes counts, $N_{1,2}$ and $N_{1,3}$ are the two-fold coincidence counts, while $N_{1,23}$ represents the three-fold coincidence counts. For this case nonclassical states are occurring for $g_c^{(2)}(0) < 1$. The same optical configuration has been used to characterize the uncorrelated thermal photons with $g^{(2)}(0) \approx 2$.

In Fig. 4 we provide extended data on the temporal waveforms of the Airy biphotons and the corresponding self-correlations $g_c^{(2)}(\tau)$ as presented in Fig. 2b of the main text and measured with the configurations of Figs. 3a and Fig. 3 b respectively.

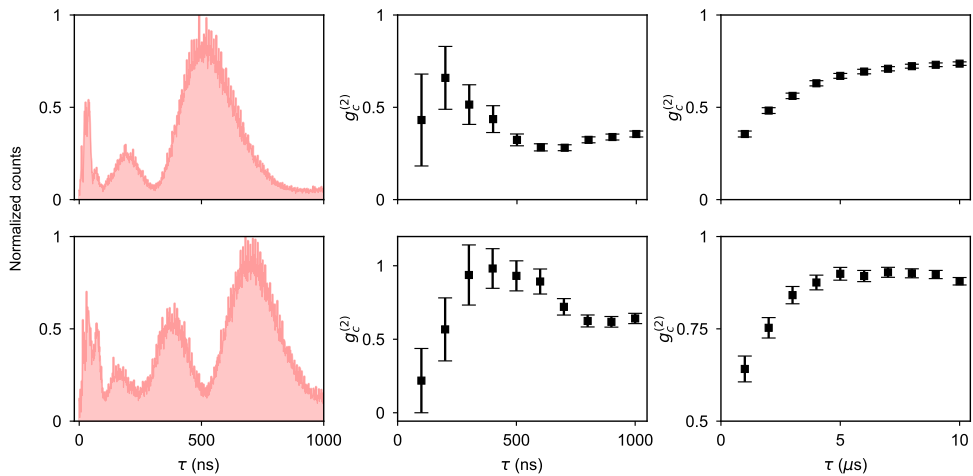

**Fig. 4 | Nonclassical correlations of the Airy biphotons. a,** Two-lobe Airy shaped anti-Stokes/Stokes coincidence as a function of time delay $\tau$. Self-correlation $g_c^{(2)}$ of the anti-Stokes photons heralded by the Stokes trigger **b,** for the first microsecnd, and **c,** for a longer time window extending far beyond the biphoton coherence time. **d-f,** The same as in (a-c) for a three-lobe Airy waveform.

## S.5  Conditions for the space-to-time mapping

Here we validate that the temporal biphoton waveforms $\psi(\tau)$ reported in our work can be well approximated, as proposed by [1, 2], by the Eqn. 2 of the main text:

$$\psi\left(\tau\right) \propto \chi^{(3)}\left(0\right)E_cE_pV_gf_p\left(\frac{L}{2}-V_g\tau\right),\tag{2}$$

Specifically, the main relevant physical conditions satisfied throughout our Airy biphoton experiments are: (I) An undepleted, modulated, weak, far-detuned pump beam that is clearly satisfied since the pump beam is Airy shaped, with a Rabi frequency $\Omega_p \approx 2\pi \times 1\,\mathrm{MHz}$ with a detuning of $\Delta_p = 120\,\mathrm{MHz}$. (II) An atomic cloud and a coupling beam with longitudinally uniform profiles, that is approximately the case since the $3.0°$ inclined pumping beam is illuminating around $1\,\mathrm{mm}$ of the $L = 1.8\,\mathrm{cm}$ atomic ensemble and the coupling beam has a wide Gaussian profile with an $1/e^2$ confinement of $12\,\mathrm{mm}$ along the $z-$direction. (III) $\chi_s^{(1)} \approx 0$ that is a direct consequence of condition (I) which has the implication that the Stokes photons are basically unaffected by the atomic medium, and, thus, propagating with the speed of light. (IV) A negligible photon loss along propagation, which leads to real $\Delta k \approx \omega/V_g$, ($\Delta k\left(z\right) \equiv k_{as}\left(z\right) - k_s\left(z\right) - (k_c - k_p)\cos\theta$) and is evident for the Stokes photons when (III) holds, while for the anti-Stokes photons it is the case when the coupling Rabi frequency is sufficiently large. For $OD$ values of around 150 and $\Omega_c \approx 2\pi \times 10\,\mathrm{MHz}$ the transmission of the photons is larger than 95%. We note that for the specific case of Airy biphotons $Ai\left(\tau/\tau_0\right)\exp\left(a\tau\right)$ any losses will only modify the exponential containment factor $a$. And, finally, (V) $\chi^{(3)}\left(\omega\right) \approx \chi^{(3)}\left(0\right)$ for a nonlinear susceptibility that is spectrally much broader than the sub-MHz EIT window $\Delta\omega_{EIT} \approx |\Omega_c|^2/2\gamma_{13}\sqrt{OD}$ for optical depths $OD \geq 4\pi^2$ that are sufficient for multi-pulse biphoton generation.

## S.6  Waveform dependence on the coupling power and the optical depth

The space-to-time mapping as expressed in Eqn. 2 for the same coupling power range and optical depth used in Figs. 3b, d and f of the main text is shown in Figs. 5a-c respectively. The agreement with the observed linearly shifting waveform is evident while the main difference is that - as expected - no precursor photons are predicted by the analytical expression of Eqn. 2.

Below we provide additional numerical simulations and analytical results based on the pump beam profile $f_p\left(z\right)$ to complement Fig. 2c of the main text. The temporal waveform for a multi-lobe Airy beam is plotted here as a function of the optical depth and it is demonstrated that indeed higher optical densities can lead to quantum Airy biphotons with more oscillations.

## S.7  Waveform dependence on the pump shape

The propagation angle $\theta = 3.0°$ of the classical pumping beam scales the horizontal shape of the 1D-Airy beam by a factor of $1/\sin\left(\theta\right) = 18.7$ and leads to the longitudinal distribution $f_p\left(z\right)$. To illustrate the effect of the $f_p\left(z\right)$ we have performed additional simulations for the case that the pump size is changed, Figs. 7a-c and its position with respect to the MOT center

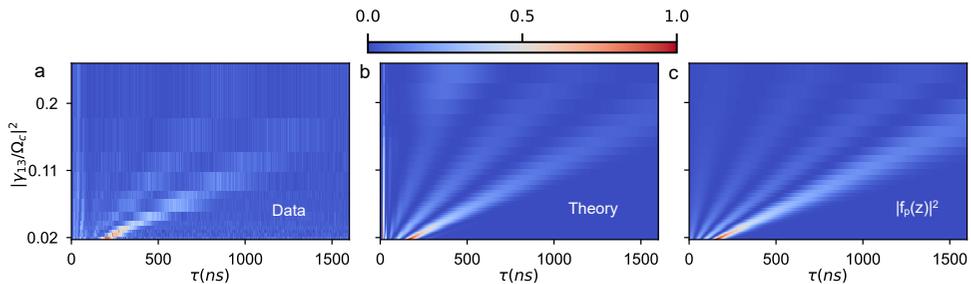

**Fig. 5 | Temporal Airy waveform dependence on the coupling power.** Temporal waveform density $|\psi(\tau)|^2$ as a function of $1/|\Omega_c|^2$. **a,** Experiment from Fig. 3 of the main text, **b,** the corresponding simulation, and **c,** the analytical expression based on Eqn. 1 of the main text. The linear dependence on $1/|\Omega_c|^2$ is evident.

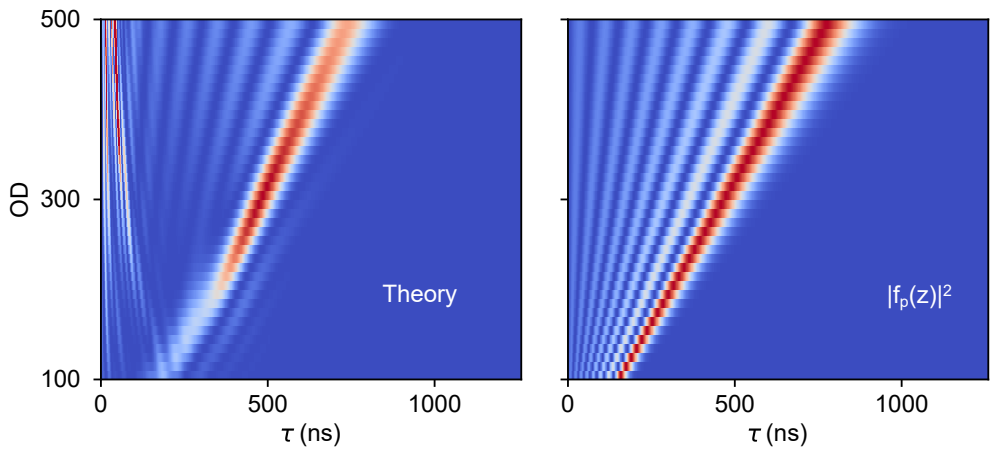

**Fig. 6 | Temporal Airy waveform dependence on the optical depth.** Biphoton waveforms for a fixed coupling Rabi frequency of $|\Omega_c| = 6\gamma_{13}$ as in Fig. 2c of the main text. Theoretical plots based on **a,** numerical simulations from the interaction picture, and **b,** the analytical form $|f_p(z)|^2$.

is shifted Figs. 7d-f. These effects have been also encountered while we were experimentally optimizing the temporal Airy waveform. When the pump Airy beam is larger than the atomic cloud, the temporal waveform will be truncated, whereas, if the Airy beam has very small features, the finite optical depth will limit the number of the observed lobes. Shifting the position of the pump beam too far from the center of the MOT, either towards the anti-Stokes or the Stokes side, will affect the temporal features due to truncation of or unwanted slow light inhomogeneities.

In Fig. 8 we present additional evidence on the effect of the pump phase on the waveform of the generated biphotons. Instead of an $f_p(z) = Ai(x/x_0)\exp(ax/x_0)$, which has alternating 0 and $\pi$ phases between its lobes, a flat phase $|f_p(z)|$ is simulated and compared with the observed temporal profiles. It is evident that this flat phase pattern cannot reproduce our experimental observations and the local minima do not drop to zero anymore. This

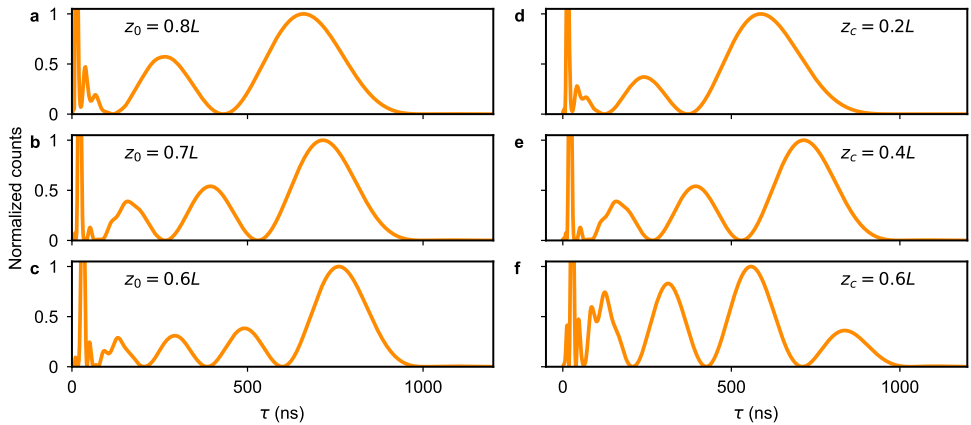

**Fig. 7 | Temporal Airy waveform dependence on the pump shape** $f_p(z)$. Normalized counts as a function of the delay time $\tau$ for three different **a-c,** pump beam size parameters $z_0$, and **d-f,** locations $z_c$ of the Airy pump beam maximum with respect to the MOT center.

is in agreement with Eqn. 2, which indicates that the quantity determining the biphoton shape is the electric field, and not the intensity of the pump beam. The discontinuous derivative of the electric field on the points where the phase flips could be the explanation of this dissimilarity when the atomic medium is not very dense.

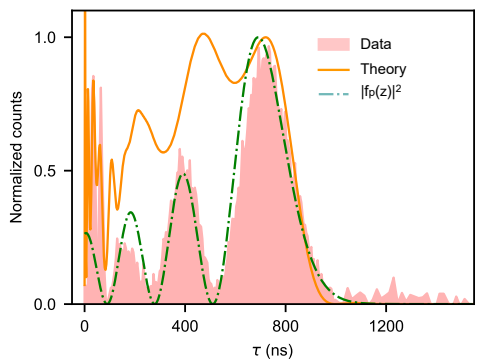

**Fig. 8 | Temporal waveform of a flat-phase Airy pump.** Experimental data (pink shaded area) is compared with the analytical $|f_p(z)|^2$ (green dotted line), and the simulated waveform (orange line) for the same conditions as in Fig. 2a of the main text.

## S.8 Measurement of the spatiotemporal single photon Airy wave packet

Because of the separability of the three-dimensional free-particle Schrödinger-like equation, which governs spatiotemporal dynamics in homogeneous dispersive media, the waveform of the temporal wavapacket can be measured in any

position of the spatial distribution, or even in a large-area detector, which collects all the incident light. To maximize the detected photon rates we have first measured the temporal waveform by collecting the photons before spatial modulation. To verify the validity of this approach we have also measured and compared the temporal waveforms on the maxima of the first and second lobes of the single-photon spatial Airy wave packets, as well as the waveform before spatial modulation by the SLM. As seen in Fig. 9 the normalized temporal profiles are, as expected, practically identical.

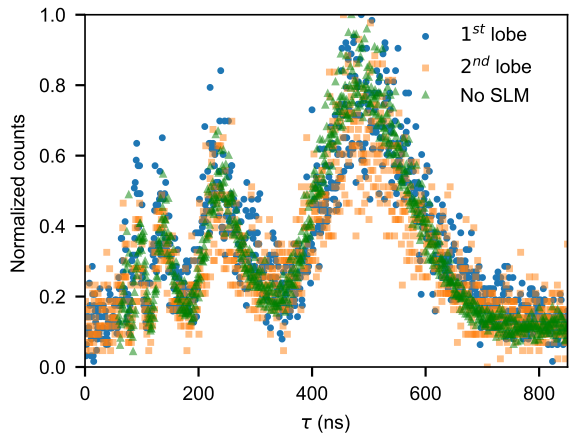

**Fig. 9  | Temporal waveform dependence on photon collection location.** Measured temporal waveform at the peak of the first (blue dots), the second (orange dots) lobe along $x$ of the two-dimensional Airy photon profile, as well as (green dots) directly after the anti-Stokes collection fiber without passing from the beamshaping optics.

## S.9   Spatial phase modulation of single photons

The essential components of the optical setup for the generation and collection of the Airy-shaped single photons are an SLM, a Fourier-transforming lens, and a collecting fiber as Fig. 1a of the main text shows. $(x', y')$ and $(x, y)$ represent the coordinates on the SLM plane and the observation plane (at the fiber input) respectively. The incident photons can have an Airy temporal waveform while on their spatial profile, which is initially Gaussian, a superposition of linear, quadratic, and cubic phases is imprinted. The cubic phase determines the intensity features of the single-photon Airy beam, while the linear and quadratic phases provide a precise and convenient way to spatially scan the three-dimensional Airy photons along the transverse plane $xy$, and emulate the propagation along the $z$ direction. To quantify this programmable spatial scanning we express the phase distribution as follows:

$$\sum_{n=1,2,3} \left( \Delta_{n,x'} x'^n + \Delta_{n,y'} y'^n \right), \tag{3}$$

where the transverse coordinates on the SLM plane are $x' = i \cdot d$ and $y' = j \cdot d$, and $i, j$ are the corresponding SLM pixel indices.

The cubic phase terms $\Delta_{3,x'}x'^3 + \Delta_{3,y'}y'^3$ contribute to the generation of Airy beam, and for simplicity we set $\Delta_{3,x'} = \Delta_{3,y'} = \Delta_3$ to keep the two axes symmetric. We consider an incident round Gaussian beam:

$$U(x', y') = U_0 \exp\left(-\frac{x'^2 + y'^2}{w_0^2}\right). \tag{4}$$

The beam amplitude on the collecting plane, $z = 0$, can be expressed [3] by the Fourier transform of the product of the incident Gaussian beam and the cubic phase as $U_0 \exp\left[-\left(x'^2 + y'^2\right)/w_0^2\right]\exp\left[i\Delta_3(x'^2 + y'^2)\right]$. The output amplitude is given by the following expression:

$$U(x, y) \propto Ai\left(\frac{x}{x_0}\right) Ai\left(\frac{y}{x_0}\right)\exp\left(-a\frac{x+y}{x_0}\right), \tag{5}$$

where $x_0$ is the scale factor for both $x$ and $y$ axes, while $a$ is the confinement factor. The scale factor determines the size of the Airy lobes and the confinement factor the characteristic Airy beam decay. The scale and confinement factors are explicitly given by the following expressions [4]:

$$x_0 = \frac{(3\Delta_3)^{1/3}\lambda f}{2\pi}, \tag{6}$$

and

$$a = \frac{1}{w_0^2 (3\Delta_3)^{2/3}}. \tag{7}$$

Typical experimental parameters $\Delta_3 = 1400^3\text{m}^{-3}$, $\lambda = 795\,\text{nm}$ and $f = 400\,\text{mm}$, lead to an Airy beam with $x_0 = 102\,\mu\text{m}$, $a = 0.303$.

The quadratic phase is applied to emulate the propagation of the beam field. The relation between the coefficients $\Delta_{2,x}$, $\Delta_{2,y}$ and the distance $z$ can be demonstrated under the Fresnel approximation [5]. For the sake of simplicity, we set $\Delta_{2,x} = \Delta_{2,y} = \Delta_2$. The phase of the beam is modulated by the quadratic term $\Delta_2\left(x^2 + y^2\right)$ on the SLM placed at the back focal plane of the lens. The scalar electric field at the front focal plane can be expressed as follows:

$$\begin{aligned}
U(x, y) &= \mathcal{F}\left[U(\xi, \eta)\exp\left[-i\Delta_2\frac{\lambda^2 f^2}{4\pi^2}\left(\xi^2 + \eta^2\right)\right]\right]\\
&= \mathcal{F}\left[U(\xi, \eta)\right] * \mathcal{F}\left[\exp\left[-i\Delta_2\frac{\lambda^2 f^2}{4\pi^2}\left(\xi^2 + \eta^2\right)\right]\right],
\end{aligned} \tag{8}$$

where $U(x, y)$ and $U(\xi, \eta)$ denote the complex amplitude at distance $z$ and $z = 0$ respectively, while $\xi = \frac{x}{\lambda f}$, $\eta = \frac{y}{\lambda f}$ are the scaled spatial coordinates on the $z = 0$ plane.

$$\mathcal{F}\left[\exp[-i\Delta_2\frac{\lambda^2 f^2}{4\pi^2}(x^2 + y^2)]\right] \propto \exp\left[-\frac{i\pi^2}{\Delta_2\lambda^2 f^2}\left(x^2 + y^2\right)\right]. \tag{9}$$

We give below the quantitative relation between the quadratic phase modulation and the equivalent free-space propagation of light along $z$ for our experiments. Without SLM modulation, the phase on back focal point of lens is $\mathcal{F}\{U(\xi,\eta)\}$. The electric field envelope of the generated beam can be described by the convolution:

$$U(x,y) = \iint_{-\infty}^{+\infty} \mathcal{F}[U(\xi,\eta)]\, h(x-\xi, y-\eta)\, d\xi d\eta = \mathcal{F}[U(\xi,\eta)] * h(x,y). \quad (10)$$

The convolution kernel $h$ is [5]:

$$h(x,y) = \frac{e^{ikz}}{i\lambda z} \exp\left[\frac{ik}{2z}\left(x^2 + y^2\right)\right]. \quad (11)$$

By directly comparing Eqn. 8 and Eqn. 10, the effective propagation distance $z$ as a function of the quadratic phase $\Delta_2$ is:

$$z = \frac{\lambda}{\pi} f^2 \Delta_2. \quad (12)$$

For our experiments, the change of the quadratic phase $\Delta_2$ leads to a total propagation distance of approximately $0.8\,\mathrm{m}$ with an interval between each measurement of $16\,\mathrm{mm}$.

The linear phase $\Delta_{1,x}x + \Delta_{1,y}y$ is applied to shift the pattern generated on the fiber collecting plane along the $x$ and $y$ directions by $l_x$ and $l_y$ respectively:

$$l_x, l_y = \frac{\lambda f \Delta_{x,y}}{2\pi}. \quad (13)$$

In our experiments, we change the $\Delta_x$ and $\Delta_y$ to move the pattern by a total of $714\,\mathrm{\mu m}$, and the interval between each measurement is $30.2\,\mathrm{\mu m}$. In practice, the shift resolution limit is set by the SLM pixel grid, and the 8-bit range of the phase modulation. For our SLMs the minimum moving step can be in the sub-micrometer scale.

## S.10  Fitting the two-dimensional Airy beam

The scalar electric field of the 2D Airy beam can be expressed as:

$$U(x,y,z=0) = U_0 Ai\left(\frac{x-x_s}{x_0}\right) Ai\left(\frac{y-y_s}{y_0}\right) \exp\left(-a_x \frac{x-x_s}{x_0} - a_y \frac{y-y_s}{y_0}\right). \quad (14)$$

The fitting parameters are $x_0$, $y_0$, $a_x$, $a_y$ for the shape, the peak amplitude $U_0$, and the position shifts $x_s$ and $y_s$. The single photon Airy beam is fitted to the intensity distribution $|U(x,y,z=0)|^2$ using the least-squares method. Here we present the data that correspond to the maximum of the $\alpha/\beta$ curve of Fig. 5b of the main text. The retrieved fitting parameters are listed in Table 1

and the comparison of the experimental measurement and the corresponding fitting is shown in Fig.10.

| $a_x = a_y$ | $x_0 = y_0$ (mm) | $I_0$ | $x_s$ (mm) | $y_s$ (mm) |
| --- | --- | --- | --- | --- |
| 0.141 | 0.938 | 151.7 | 0.534 | 0.554 |

**Table 1** Fitting parameters

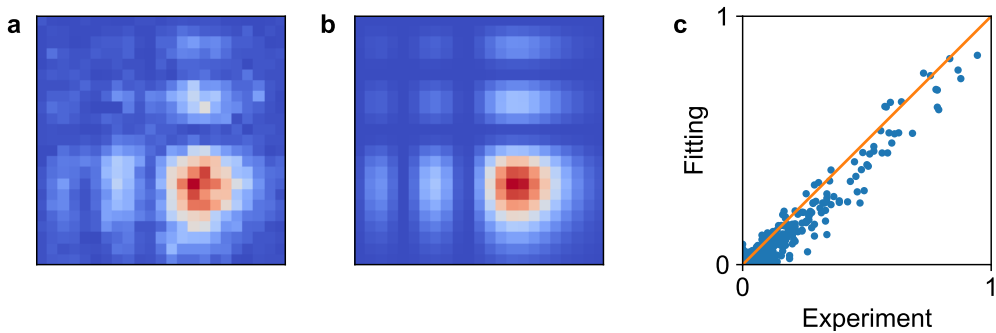

**Fig. 10 | Fitting the Airy-shaped single photons in the presence of noise. a,** Experimental Airy-shaped single photon with superimposed noise, and **b,** fitting result of the Airy-shaped photon. **c,** Linear regression comparison of the experimental fitting results. The slight deviation from the y=x line is explained by the always additive influence of the weak noise light.

## S.11 Concealing correlated Airy photons in classical light noise

The anti-Stokes photons are modulated in space as 2D Airy wave packets and external classical noise is added, as shown in Fig. 1a of the main text. The total collected signal Airy+noise can be decomposed using the natural basis choice of Airy and noise patterns as presented below:

$$N_{total}\,(x,y,t_w) = \alpha\,(t_w)\,N_{Airy}\,(x,y) + \beta\,(t_w)\,N_{noise}\,(x,y)\,, \qquad (15)$$

where $N_{total}$, $N_{Airy}$ and $N_{noise}$ are the normalized spatial distributions for all collected photons, and the basis of Airy beam and noise respectively when $\Sigma_{x,y}N_{total} = \Sigma_{x,y}N_{Airy} = \Sigma_{x,y}N_{noise} = 1$. The total photon distribution $N_{total}$ is normalized within a certain time window $t_w$. $\alpha$ and $\beta$ are the projection coefficients for the relative portion of the Airy and noise photons that satisfy $\alpha + \beta = 1$.

The Airy basis is retrieved by the normalized fitting result of section S.10 as shown in Fig. 11a, and, similarly, the basis pattern for the noise is extracted by the normalized collected spatial data only in the presence of noise as shown in Fig. 11b. $\alpha$ and $\beta$ are retrieved as a function of the time windows $t_w$. Typical experimental data and the corresponding fitting results are shown in Fig. 11c and d.

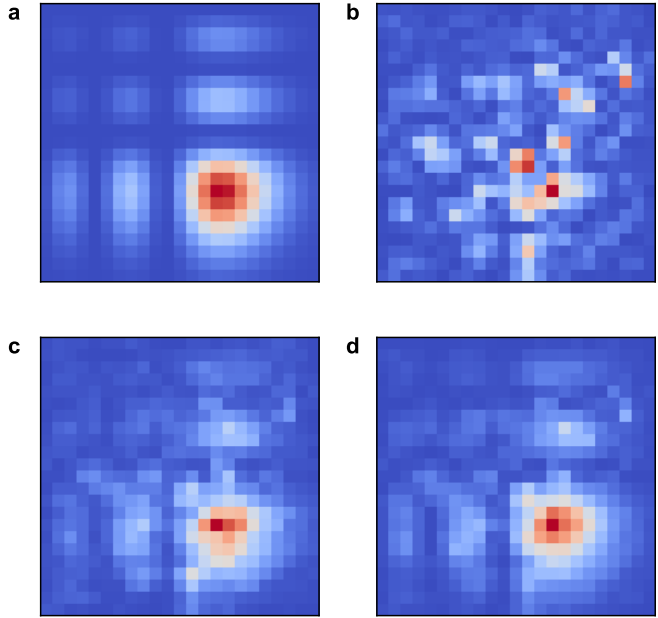

**Fig. 11 | Spatial decomposition of the observed photon distributions.** Decomposition basis **a,** 2D Airy, and **b,** noise. **c,** Experimental data of the collected Airy+noise. **d,** Fitting result with the projection on the Airy and noise basis functions.

We have developed a simple model to predict the time evolution of the projection coefficients $\alpha$ and $\beta$ when the temporal waveform of the biphotons is known. The Stokes and anti-Stokes photons are generated in a steady rate during the 1 ms of the SFWM measurement window (Fig. 1). The Stokes photons travel at the speed of light, while the speed of the slow anti-Stokes photons is determined by the EIT condition of the atomic medium. Therefore, with the Stokes photons acting as trigger, the time delay of the correlated anti-Stokes photons is not equally distributed in time as Fig. 12a shows. $R_{Airy}(t)$ and $R_{noise}(t)$ present the generation rate for a Stokes trigger, and thus, $\alpha$ and $\beta$ are given by:

$$\alpha(t_w) = \frac{\int_0^{t_w} R_{Airy}(t)\mathrm{d}t}{\int_0^{t_w} R_{Airy}(t)\mathrm{d}t + \int_0^{t_w} R_{noise}(t)\mathrm{d}t} \tag{16}$$

$$\beta(t_w) = \frac{\int_0^{t_w} R_{noise}(t)\mathrm{d}t}{\int_0^{t_w} R_{Airy}(t)\mathrm{d}t + \int_0^{t_w} R_{noise}(t)\mathrm{d}t}, \tag{17}$$

where $R_{noise}$ can be treated as constant while the waveform of the $R_{Airy}$ within the photon coherence time is engineered by the profile of the pump, $f_p$. For a Gaussian pump, $R_{Airy}(t)$ has a typical waveform as the own shown in Fig. 12a. The theoretically predicted ratio $\alpha(t_w)/\beta(t_w)$ is plotted in Fig. 12b and it is in agreement with our observations from Fig. 5b of the main text.

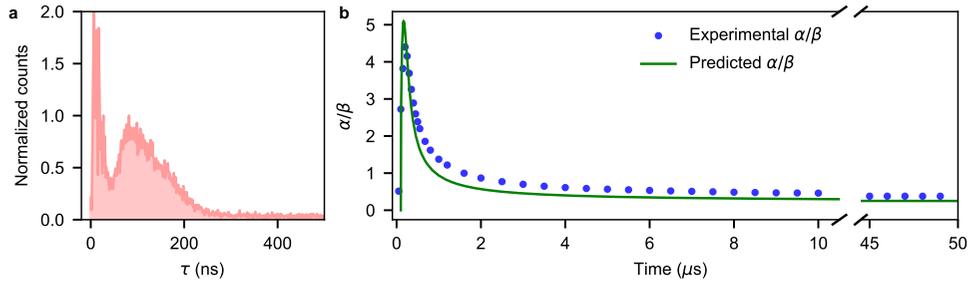

**Fig. 12 | Time evolution of the spatial decomposition. a,** Typical biphoton shape for a Gaussian pump beam. **b,** Time dependence of the experimental (blue dots) and modeled (solid green line) ratio $\alpha/\beta$ between the Airy, $\alpha$ and noise, $\beta$ projection coefficients.



\end{document}